\documentclass[onecolumn,showpacs,nofootinbib,superscriptaddress,aps,prd,11pt]{revtex4}
\usepackage{graphicx,dcolumn,bm,amsfonts,amsmath,color,xcolor,amsthm}
\usepackage[colorlinks=true, citecolor=blue, linkcolor=blue,urlcolor=blue]{hyperref}
\usepackage{slashed}
\usepackage{tabularx}
\usepackage{ulem}
\allowdisplaybreaks

\begin{document}
\title{$B_{(s)} \to S(a_0(1450), K_0^*(1430), f_0(1500))$ helicity form factors within QCD light-cone sum rules}

\author{Yi Zhang}
\affiliation{School of Physics, Beihang University, Beijing 102206, China\\
Centrale Pekin, Beihang University, Beijing 100191, China\\
Peng Huanwu Collaborative Center for Research and Education, Beihang University, Beijing 100191, China\\
Beijing Key Laboratory of Advanced Nuclear Materials and Physics, Beihang University, Beijing 102206, China\\
Southern Center for Nuclear-Science Theory (SCNT), Institute of Modern Physics, Chinese Academy of Sciences, Huizhou 516000, China}

\author{Wei Cheng}
\affiliation{School of Science, Chongqing University of Posts and Telecommunications, Chongqing 400065, China}

\author{Jia-Wei Zhang}
\affiliation{Department of Physics, Chongqing University of Science and Technology, Chongqing 401331, China}

\author{Tao Zhong}
\affiliation{Department of Physics, Guizhou Minzu University, Guiyang 550025, China}
\affiliation{Institute of High Energy Physics, Chinese Academy of Sciences, Beijing 100049, China}

\author{Hai-Bing Fu}
\email{fuhb@gzmu.edu.cn}
\affiliation{Department of Physics, Guizhou Minzu University, Guiyang 550025, China}
\affiliation{Institute of High Energy Physics, Chinese Academy of Sciences, Beijing 100049, China}

\author{Li-Sheng Geng}
\email{lisheng.geng@buaa.edu.cn}
\affiliation{School of Physics, Beihang University, Beijing 102206, China\\
Centrale Pekin, Beihang University, Beijing 100191, China\\
Peng Huanwu Collaborative Center for Research and Education, Beihang University, Beijing 100191, China\\
Beijing Key Laboratory of Advanced Nuclear Materials and Physics, Beihang University, Beijing 102206, China\\
Southern Center for Nuclear-Science Theory (SCNT), Institute of Modern Physics, Chinese Academy of Sciences, Huizhou 516000, China}

\begin{abstract}
In this paper, we investigate the helicity form factors (HFFs) of the $B_{(s)}$-meson decay into a scalar meson with a mass larger than 1~GeV, {\it i.e.,}  $B \to a_0(1450)$,  $B_{(s)} \to K_0^*(1430)$ and $B_{s} \to f_0(1500)$  by using light-cone sum rules approach. We take the standard currents for correlation functions. To enhance the precision of our calculations, we incorporate the next-to-leading order (NLO) corrections and retain the scalar meson twist-3 light-cone distribution amplitudes. Furthermore, we extend the HFFs to the entire physical $q^2$ region employing a simplified $z$-series expansion. At the point of $q^2=1\rm{~GeV^2}$, all NLO contributions to the HFFs are negative, with the maximum contribution around $25\%$. Then, as applications of these HFFs, we analyze the differential decay widths, branching ratios, and lepton polarization asymmetries for the semi-leptonic $B_{(s)} \to S \ell \bar{\nu}_\ell$, FCNC $B_{(s)} \to S \ell \bar{\ell}$ and rare $B_{(s)} \to S \nu \bar{\nu}$ decays. Our results are consistent with existing studies within uncertainties. The current data still suffer from large uncertainties and need to be measured more precisely, which can lead to a better understanding of the fundamental properties of light scalar mesons.
\end{abstract}


\maketitle

\section{Introduction}
The heavy mesons weak decays have attracted a lot of attention to the theoretical and experimental groups, especially the bottom meson with mass larger than 1 GeV, {\it i.e.} $B_{(s)} \to S$ with $S$ denotes $a_0(1450), K_0^*(1430)$ and  $f_0(1500)$ meson~\cite{Yang:2005bv,Huang:2024uuw,Lu:2018cfc}, due to the significance understanding the fundamental properties for hadronic physics of these processes. The significance stems from their unique capacity to serve as a testing ground for multiple fundamental aspects of particle physics: precision tests of the Standard Model (SM), determination of Cabibbo-Kobayashi-Maskawa (CKM) matrix elements, investigation of CP-violation phenomena, and evaluation of both perturbative and nonperturbative quantum chromodynamics (QCD) methodologies. On the other hand, they are the key way to investigate scalar mesons fundamental structures and properties within the SM and search for new physics beyond the SM. Therefore, the semi-leptonic decays of $B_{(s)}$-mesons to light scalar mesons, especially those near 1.5 GeV, have attracted much attention.

In recent years, the Belle/BelleII, BaBar, LHCb collaborations have obtained many remarkable results~\cite{Belle-II:2021rof, CLEO:1993nic, Nicolini:2024dar, CLEO:1994veu, Belle:2001oey, BaBar:2003szi, LHCb:2022vje, CDF:2011tds, CDF:2011grz,  Belle:2009zue, Belle:2005hjc, Belle:2008knz, BaBar:2008fao, LHCb:2015svh, LHCb:2011tcp}. Various decay modes of the $B_{(s)}$-mesons to scalar mesons have been observed and accurately measured, including the $B_{(s)}$-mesons decay into scalar meson flavor changing charged current~(FCCC) semileptonic decays and flavor changing neutral current~(FCNC) decays. However, the FCCC semileptonic decays $\bar{B}^0 \to a_0 (1450)^+ \ell \bar{\nu}_\ell$ and FCNC decays $B_s \to f_0(1500) \ell \bar\ell$ occurring through the transition $b \to u \ell \bar\nu_\ell$, $b \to s\ell\bar\ell$ at the quark level, have not been observed yet. The internal structure of scalar mesons - whether they exist as conventional two-quark states ($q\bar{q}$), tetraquark configurations ($qq \bar q \bar q$), or molecular components - remains an ongoing subject of debate in hadron physics. According to the experimental data and the quark model, the scalar mesons observed below and above $1~\rm{GeV}$ can be arranged into two SU(3) nonets~\cite{Cheng:2005nb, Lu:2006fr, Du:2004ki}. There are two competing scenarios for these scalar mesons. In scenario $1$, the light scalar mesons are assumed to be composed of two quarks. The nonet mesons below $1 ~\rm{GeV}$ are treated as the lowest-lying states, and the nonet mesons near $1.5~\rm{GeV}$ are the excited states. In scenario 2, those scalar mesons below $1~\rm{GeV}$ may be considered as members of four-quark bound states~\cite{Jaffe:1976ig}, meson-meson molecular states~\cite{Lee:2022jjn, Weinstein:1982gc}, tetraquark states~\cite{Jaffe:1976ih,Wang:2010pn}, glueballs~\cite{Amsler:2004ps}, and the superposition of these states~\cite{Amsler:1995tu}; the other nonet mesons are composed of two quarks and viewed as the lowest-lying states. Notably, within the LCSR framework, a systematic investigation of helicity form factors for $B_{(s)} \to S$ transitions (where $S = a_0(1450), K_0^*(1430), f_0(1500)$) under the two-quark picture remains absent in the literature. Given these theoretical uncertainties, our study adopts the conventional quark-antiquark paradigm as its foundational assumption, aiming to provide new theoretical benchmarks for experimental investigations of scalar meson structures. In this work, we follow the latter scenario.

In the theoretical studies of the $B_{(s)} \to S$ decays, one of the most challenging parts are the precise calculation of those form factors. The dominant contribution to these form factors arises from short-distance dynamics in the large recoil region. They have been studied in various approaches, such as perturbative QCD (pQCD) \cite{Li:2008tk}, QCD sum rules (QCDSRs)~\cite{Ghahramany:2009zz, Han:2013zg}, light-cone sum rules (LCSRs)~\cite{Han:2023pgf, Huang:2022xny, Wang:2014vra, Sun:2010nv}, and the relativistic quark model (RQM)~\cite{Faustov:2013ima}, etc. It should be noted that these approaches are typically limited to different $q^2$-regions. The corresponding results must be extrapolated to the whole physical region by adopting appropriate parametrizations. Generally, the LQCD performs well in the small-recoil region, while the pQCD factorization is applicable near the large recoil point. Estimations based on the QCDSRs are effective in lower and intermediate $q^2$-regions, respectively. Conversely, the TFFs calculated with LCSRs are parameterized as the initial or final meson light-cone distribution amplitudes (LCDAs), organized into various twist structures. Starting from different correlation functions, the LCSRs of $B \to S$ TFFs are expressed as convolution integrals of the scalar meson\cite{Huang:2022xny} or $B$-meson LCDAs~\cite{Khosravi:2022fzo}.

In the present work, we shall adopt the LCSRs approach to calculate the HFFs of the $B_{(s)}$ mesons decay into the light scalar meson $a_0(1450)$, $K_0^*(1430)$, $f_0(1500)$, {\it i.e. $B_{(s)} \to S$} by adopting the conventional correlation functions. Different from the previous LCSRs~\cite{Zhong:2022ecl,Zhong:2021epq}, we shall express the hadronic matrix elements using the HFFs with the help of the covariant helicity projection approach~\cite{Korner:1989qb}. Theoretically, one generally parameterizes these hadron matrix elements as form factors, such as $f_+{(q^2)}$, $f_0{(q^2)}$ and $f_T{(q^2)}$, which play a crucial role in describing the $B_{(s)} \to S$ transitions. In recent years, a new theoretical method called helicity form factors (HFFs) has emerged. These include the longitudinal helicity form factor $\mathcal{S}_0(q^2)$ and transverse helicity form factors $\mathcal{S}_\pm(q^2)$. Compared to the usual transition form factors, HFFs are more natural, simplifying theoretical calculations and improving the interpretation of experimental data in some cases. In addition, the dispersive bounds on the HFFs parametrization can be achieved via the diagonalizable unitarity relations. The LCSRs for the HFFs can be conveniently used to study the physical observables that encode the dynamical information. More information can be found in Refs.~\cite{Cheng:2022mvd,Bharucha:2010im, Cheng:2018ouz, Cheng:2021svx, Fu:2020uzy}. It is well known that the LCSRs show strong computational power and theoretical prediction ability in dealing with hadron physical processes. According to the standard LCSRs method, we calculate the HFFs of the $B_{(s)} \to S$ transitions from the conventional correlation function so the contributions of the different twist distribution amplitudes to the HFFs can be taken into consideration, the next-to-leading order contributions are also taken into account. After calculating the HFFs of the $B_{(s)} \to S$ transitions within the LCSRs, we extrapolate them into the allowed physical region and accurately analyze the differential decay width and branching ratio of the FCCC semileptonic decays $B_{(s)} \to S\ell\bar{\nu}$, the FCNC semileptonic decays $B_{(s)} \to S \ell^+\ell^-$ and the rare decay $B_{(s)} \to S\nu\bar{\nu}$, taking into account the influence of the lepton mass.

The remaining parts of the paper are organized as follows. In the following Sec.~\ref{2}, we present the calculation technology for the HFFs of the $B_{(s)} \to S$ transitions within the LCSRs approach. In Sec.~\ref{3}, we provide a detailed numerical discussion. By extrapolating those HFFs to the whole physical $q^2$-region, we rigorously analyze the form factors and differential widths of the FCCC semileptonic decays $B_{(s)} \to S\ell\bar{\nu}$, FCNC semileptonic decays  $B_{(s)} \to S \ell^+\ell^-$ and rare decay $B_{(s)} \to S\nu\bar{\nu}$, including a comparison with the prediction of some other theoretical approaches. The last section is reserved for a summary.

\section{Computing the $B_{(s)} \to S$ HFFs}\label{2}

To derive the HFFs of the $B_{(s)} \to S$ transitions, one can project out the relevant HFFs from the vector and tensor hadronic matrix elements by using the off-shell $W$-boson polarization vector, which can be written as follows~\cite{Bharucha:2010im}:
\begin{align}
&\mathcal{S}_{V,\sigma}(q^2 ) \! = \! \sqrt{\frac{q^2}{\lambda}} {\epsilon_\sigma^{*\mu}(q)} \langle S(k) |\bar q  \gamma_\mu \gamma_5  b |B_{(s)}(p)\rangle,\\
&\mathcal{S}_{T,\sigma}(q^2 )  \!=\! (-i)\sqrt{\frac{1}{\lambda}}  {\epsilon_\sigma^{*\mu}(q)} \langle S(k) |\bar q  \sigma_{\mu\nu} \gamma_5 q^\nu b |B_{(s)}(p)\rangle,
\label{HFF:Definition}
\end{align}
where $q=(p-k)$ is the transfer momentum, $\lambda = (t_{-} - q^2)(t_{+} - q^2)$ with $t_\pm=(m_{B_{(s)}}\pm m_S)^2$ is a standard kinematic function. $\epsilon_\sigma^{*\mu}(q)$ are the transverse ($\sigma=\pm$), longitudinal ($\sigma=0$) or time-like ($\sigma=t$) polarization vectors, respectively. More specifically,
\begin{align}
\epsilon _ \pm ^\mu(q) &=  \mp \frac{1}{{\sqrt 2 }}  (0,1, \mp i,0), \\
\epsilon _0^\mu  (q) &=  \frac{1}{{\sqrt {q^2 } }}  (|\vec q| ,0,0, - q^0 ),  \\
\epsilon _t^\mu  (q) &= \frac{1}{{\sqrt {q^2 } }}  q^\mu .
\end{align}
Here the transverse projections will vanish in dealing with the vector form factors, where only non-zero tensor form factor is $\mathcal{S}_{T,0}(q^2)$.

In order to derive the full analytical expressions for the $B_{(s)} \to S$ HFFs within LCSR approach, one can consider the following two-point correlation functions based on the standard procedure of LCSR. The two types of correlation functions can be written as:
\begin{align}
\Pi_{V,\sigma}(p,q) &= - i \sqrt{\frac{q^2}{\lambda}}  {\epsilon_\sigma^{*\mu}(q)}\int d^4 x e^{iq\cdot x}  \langle S(p)|T\{j_{A}^\mu(x),j_{B_{(s)}}^\dag (0)\}|0\rangle ,
\label{correlator1}
\\
\Pi_{T,\sigma}(p,q) &= - i \sqrt{\frac{1}{\lambda}}  {\epsilon_\sigma^{*\mu}(q)}\int d^4 x e^{iq\cdot x}  \langle S(p)|T\{j_{T}^\mu(x),j_{B_{(s)}}^\dag (0)\}|0\rangle ,
\label{correlator2}
\end{align}
where the three types of currents have the forms
\begin{align}
&j_{A}^\mu(x) = \bar q_1 (x){\gamma _\mu } \gamma _5 b(x),\nonumber\\
&j_{T}^\mu(x) = \bar q_1 (x) \sigma _{\mu\nu} \gamma _5 q^\nu b(x), \\
&j_{B_{(s)}}^\dag (0)=i m_b \bar b(0) \gamma_5 q_2(0),\nonumber
\end{align}
which have the same quantum state of the $B_{(s)}$-meson with $J^{P}=0^-$. In Eqs.~\eqref{correlator1} and ~\eqref{correlator2}, the light quark $q_1 = u, q_2 = d$ is for $B \to a_0(1450)$ transition; the light quark $q_1 = q_2 = s$ is for the  $B_s \to f_0(1500)$ transition; the light quark $q_1 = d, q_2 = s$ is for $B_{(s)} \to K_0^*(1430)$ transition, respectively.

The hadronic representations for the two correlation functions Eqs.~\eqref{correlator1} and \eqref{correlator2} can be achieved by inserting a complete set of intermediate hadron states with the same quantum numbers as the current operator $i m_b \bar b(0) \gamma_5 q_2(0)$ between the currents in the physical region. After isolating the pole term of the lowest pseudoscalar $B_{(s)}$ meson, we can obtain the hadronic representations:
\begin{align}
\Pi _{V,\sigma}^{\rm H }&=\sqrt{\frac{q^2}{\lambda}}
\epsilon_\sigma^{*\mu}(q) \bigg[\frac{\langle S(p)
| q_1  \gamma_\mu \gamma_5  b |B_{(s)}\rangle \langle B_{(s)}|\bar b i \gamma_5 q_2|0\rangle }{m_{B_{(s)}}^2 - (p + q)^2}
\nonumber
\\
& +  \sum\limits_{\rm H}  \frac{\langle S(p)
| \bar{q}_1  \gamma_\mu \gamma_5  b |B_{(s)}^H\rangle \langle B_{(s)}^H|\bar b i \gamma_5 q_2 |0\rangle }{m_{B_{(s)}^H}^2 - (p  +  q)^2}\bigg],
\\
\Pi _{T,\sigma}^{\rm H }&=\sqrt{\frac{1}{\lambda}}
 \epsilon_\sigma^{*\mu}(q) \bigg[  \frac{\langle S( p )
| \bar{q}_1 (x) \sigma _{\mu\nu} \gamma _5 q^\nu b(x) |B_{(s)}\rangle \langle B_{(s)}|\bar b i \gamma_5 q_2|0\rangle }{m_{B_{(s)}}^2 - (p  +  q)^2}
\nonumber
\\
& + \sum\limits_{\rm H} \frac{\langle S(p) | \bar q_1 (x) \sigma _{\mu\nu} \gamma _5 q^\nu b(x)|B_{(s)}^H\rangle \langle B_{(s)}^H|\bar b i \gamma_5 q_2 |0\rangle }{m_{B_{(s)}^H}^2 - (p  +  q)^2}\bigg],
\end{align}
where $\langle B_{(s)}|\bar b i \gamma_5 q|0\rangle=m_{B_{(s)}}^2 f_{B_{(s)}}/{m_b}$ with $f_{B_{(s)}}$ being the $B_{(s)}$-meson decay constant. The phenomenological representations of the correlation functions Eqs.~(\ref{correlator1}) and (\ref{correlator2}) read
\begin{eqnarray}
\Pi _{V,\sigma}^{\rm H } &=& \frac{m_{B_{(s)}}^2 f_{B_{(s)}}}{m_b[m_{B_{(s)}}^2 - (p + q)^2]} \mathcal{S}_{V,\sigma}(q^2) + \int_{s_0}^\infty  \frac{\rho_{V,\sigma}^{\rm H}}{s - (p  +  q)^2}ds,\\
\Pi _{T,\sigma}^{\rm H } &=& \frac{m_{B_{(s)}}^2 f_{B_{(s)}}}{m_b[m_{B_{(s)}}^2 - (p + q)^2]} \mathcal{S}_{T,\sigma}(q^2) + \int_{s_0}^\infty  \frac{\rho_{T,\sigma}^{\rm H}}{s - (p  +  q)^2}ds,
\end{eqnarray}
where $s_0$ stands for the continuum threshold parameter. The spectral densities $\rho^{\rm H}_{V(T),\sigma}(s)$, denoting the excited states and continuum states, can be approximated using the ansatz of quark-hadron duality, i.e., $\rho^{\rm H}_{V(T),\sigma}(s)= \rho^{\rm QCD}_{V(T),\sigma}(s)\theta (s-s_0)$~\cite{Colangelo:2000dp,Shifman:2000jv}.

One can also calculate the correlation functions Eqs.~(\ref{correlator1}) and (\ref{correlator2}) in the deep Euclidean region by contracting the $b$-quark line into the light-cone expansion of the $b$-quark propagator $-i\int \frac{d^4k}{(2 \pi)^4} e^{-ikx} \frac{\slashed{k} +m_b}{m_b^2-k^2}$. The vacuum-to-meson matrix element can be expanded in terms of the distribution amplitudes of scalar mesons with increasing twist, which can be expressed as:
\begin{align}
& \langle S(p) | \bar{q}_{2}(x)\gamma _{\mu }q_{1}(0) | 0\rangle=p_{u}\int_{0}^{1}due^{i up\cdot x }\phi_{2;S}(u,\mu ),
\\
& \langle S(p) | \bar{q}_2(x)q_1(0) | 0\rangle=m_{S}\int_{0}^{1}due^{i up\cdot x }\psi_{3;S}^{s}(u,\mu ),  \label{DAs}
\\
& \langle S(p) | \bar{q}_{2}(x)\sigma _{\mu \nu }q_1(0) | 0\rangle=-m_{S}(p_{\mu }z_{\nu }-p_{\nu }z_{\mu }) \int_{0}^{1}due^{i up\cdot x }\psi _{3;S}^{\sigma } (u,\mu),
\end{align}
Here, $\phi_{2;S}(u,\mu)$ indicate the twist-2 DA of the scalar mesons, $\psi_{3;S}^s(u,\mu)$ and $\psi_{3;S}^\sigma(u,\mu)$ are the twist-3 DAs. Based on the definition of these matrix elements, the operator product expansion (OPE) of the correlation functions Eqs.~(\ref{correlator1}) and (\ref{correlator2}) can be obtained.

The OPE of the correlation function in the deep Euclidean  region can be matched with its hadron representation in the physical region by the dispersion relation. Furthermore, applying the Borel transformation for both sides to suppress the contributions from unknown excited and continuum states, the sum rules for the heavy-to-light HFFs in terms of the scalar mesons' LCDAs can be written as:
{\begin{eqnarray}
\mathcal{S}_{V,\sigma} (q^2) &=& \frac{e^{ {m_{{B_{(s)}}}^2} / M^2}}{f_{B_{(s)}} m_{B_{(s)}}^2}   \bigg[\mathcal{S}_{V,\sigma}^{\rm{LO}} (q^2) + \frac{\alpha_s C_F}{4 \pi}\mathcal{S}_{V,\sigma}^{\rm{NLO}} (q^2)\bigg] \\
\mathcal{S}_{T,\sigma} (q^2) &=& \frac{e^{ {m_{{B_{(s)}}}^2} / M^2}}{f_{B_{(s)}} m_{B_{(s)}}^2}   \bigg[\mathcal{S}_{T,\sigma}^{\rm{LO}} (q^2) + \frac{\alpha_s C_F}{4 \pi} \mathcal{S}_{T,\sigma}^{\rm{NLO}} (q^2)\bigg]
\end{eqnarray}
The leading-order contributions of the HFFs take the form as follows:
\begin{eqnarray}
\mathcal{S}_{V,0}^{\rm{LO}} (q^2) &=& \frac{m_b  +  m_{q_1}}{f_{B_{(s)}} m_{B_{(s)}}^2} \, e^{ {m_{{B_{(s)}}}^2} / M^2}  \,  \bigg\{ \int_{u_0}^1 \,  \frac{du}{u}  \,  e^{-({m_b^2 \,+\, u\bar u m_S^2 \,- \bar u q^2})/uM^2} \bigg[- m_b \phi_{2;S}(u)~~~~ \nonumber\\
&&  + \, m_S \Big(u \psi_{3;S}^s(u) \,+ \frac1{3}\psi_{3;S}^\sigma(u)\Big) \,+\, \frac1{uM^2} \frac{m_S}{6} (m_b^2 + u m_S^2 + q^2)\psi_{3;S}^\sigma(u)  \bigg]
\nonumber\\
&& + \frac{m_S}{6}  e^{-s_0/M^2} \frac{m_b^2 - u_0 m_S^2 + q^2}{m_b^2 + u_0 m_S^2 - q^2} \psi_{3;S}^\sigma(u_0) \bigg\}
\label{HFF AV0}
\\
\mathcal{S}_{V,t}^{\rm{LO}} (q^2) &=& \frac{m_{B_{(s)}}^2 \!-\! m_S^2 -\!q^2}{\sqrt\lambda} \bigg\{ \frac{m_b \!+\! m_q}{f_{B_{(s)}} m_{B_{(s)}}^2} \! e^{ {m_{B_{(s)}}^2} / M^2} \! \bigg\{ \!\int_{u_0}^1 \frac{du}{u} e^{-({m_b^2 + u\bar u m_S^2 - \bar uq^2})/uM^2}
\nonumber\\
&&\times \bigg[- m_b \, \phi_{2;S}(u) + m_S \,  \left(u \psi_{3;S}^s(u) + \frac1{3}\psi_{3;S}^\sigma(u)\right)\, +\,  \frac1{uM^2} \frac{m_S}{6} \psi_{3;S}^\sigma(u)
\nonumber\\
&&\times  (m_b^2 \, + \, u m_S^2 \, +\,  q^2) \bigg] \, +\,  \frac{m_S}{6} ~e^{-s_0/M^2} \frac{m_b^2 \, - \, u_0 m_S^2 \, + \, q^2}{m_b^2 \, +\,  u_0 m_S^2 \, - \, q^2} ~\psi_{3;S}^\sigma(u_0)\bigg\}
\nonumber\\
&&
+ \frac{2 q^2}{\sqrt{\lambda}} \bigg\{\frac{ m_b + m_s }{f_B m_B^2} e^{ {m_{B}^2} / M^2} \int_{u_0}^1 \frac{du }{u}  e^{-(m_b^2 +u\bar u m_S^2 - \bar uq^2)/uM^2}\bigg[m_S \, \bigg(\psi_{3;S}^s(u)
\nonumber\\
&&
+ \frac1{6u} \,  \psi_{3;S}^\sigma(u) \bigg)  \, -  \, \frac1{u^2 M^2} \frac{m_S}{6} \, (m_b^2  \, +  \, u^2 m_S^2  \, - \,  q^2) \psi_{3;S}^\sigma(u) \bigg] - \frac{1}{6 u_0} m_S
\nonumber\\
&&  \psi_{3;S}^\sigma(u_0) e^{-s_0/M^2} \bigg\} \bigg\},
\label{HFF AVt}\\
\mathcal{S}_{T,0}^{\rm{LO}} (q^2) &=& \frac{\sqrt{q^2}(m_b \!+\! m_{q_1})}{m_{B_{(s)}}^2 f_{B_{(s)}}} \! e^{m_{B_{(s)}}^2/M^2} \! \bigg\{-\frac12 \int_{u_0}^1 \! \frac{du}{u} e^{- \left( {m_b^2 + u\bar um_S^2 - \bar u q^2} \right) / uM^2}  \bigg[ \phi_{2;S}(u)
\nonumber\\
&& - \frac{m_{B_{(s)}} m_S}{3u M^2} \psi_{3;S}^\sigma(u) \bigg] + \frac{m_b m_S}{6} e^{-s_0/M^2} \psi_{3;S}^\sigma(u_0) \frac1{m_b^2 + u_0^2 m_S^2 - q^2} \bigg\},
\label{HFF AT0}
\end{eqnarray}
where $u_0 = \{-(s_0 - q^2 -m_S^2) + [(s_0 - q^2 -m_S^2)^2 + 4 m_S^2 (m_b^2 - q^2)]^{1/2}\}/(2m_S^2)$. $M^2$ is the Borel parameter. Furthermore, the next-to-leading order contributions of the HFFs read:
\begin{eqnarray}
\mathcal{S}_{V,0}^{\rm{NLO}} (q^2) &=& -\frac{m_b^2}{2f_{B_{(s)}}m_{B_{(s)}}^2}\exp\bigg( \frac{m_{B_{(s)}}^2}{M^2}\bigg)  \int_{m_b^2}^{s_0}ds\rho^{\alpha_s}_{+}(s)\exp\bigg( -\frac{s}{M^2}\bigg) ,
\end{eqnarray}
\begin{eqnarray}
\mathcal{S}_{V,t}^{\rm{NLO}} (q^2) &=& -\frac{m_b^2}{f_{B_{(s)}}m_{B_{(s)}}^2}\exp\bigg( \frac{m_{B_{(s)}}^2}{M^2}\bigg) \int_{m_b^2}^{s_0}ds\rho^{\alpha_s}_{+-}(s)\exp\bigg( -\frac{s}{M^2}\bigg) ,
\end{eqnarray}
\begin{eqnarray}
\mathcal{S}_{T,0}^{\rm{NLO}} (q^2) &=& -\frac{m_{B_{(s)}}+m_S}{2f_{B_{(s)}}m_{B_{(s)}}^2}\exp\bigg( \frac{m_{B_{(s)}}^2}{M^2}\bigg) \int_{m_b^2}^{s_0}ds\rho^{\alpha_s}_{T}(s)\exp\bigg( -\frac{s}{M^2}\bigg) ,
\end{eqnarray}
For the QCD spectral densities $\rho_{+}^{\alpha_s}(s)$, $\rho_{+-}^{\alpha_s}(s)$ and $\rho_{T}^{\alpha_s}(s)$, one can refer to Ref. \cite{Wang:2014vra}.

\section{Numerical results and Discussions}\label{sec:Numerical results and discussion} \label{3}

In this part, we investigate the phenomenological applications of the HFFs. We present the HFFs, the differential decay widths, the branching ratio, and the longitudinal lepton polarization asymmetries for the FCCC semi-leptonic $B_{(s)}\to S \ell \bar{\nu}_\ell$, rare FCNC semileptonic $B_{(s)}\to S \ell \bar{\ell}$ and $B_{(s)}\to S \nu \bar{\nu}$ decays. First, we discuss the DA of the scalar mesons and analyze the choices for the Borel parameters $M^2$ and the threshold parameters $s_0$. Then, the $z$ parameterization of the form factors is used to extend the HFFs to the whole kinematically physical region. Finally, we calculate the relevant physical observables based on the HFFs.
\begin{table}[tb]
\footnotesize
\caption{Scalar meson decay constants and the Gegenbauer moments of the twist-2, 3 LCDAs $\phi_{2;S}(u,\mu)$, $\psi_{3;S}^{s(\sigma)}(u,\mu)$ at the scale
$\mu=1~{\rm{GeV}}$ \cite{Cheng:2005nb,Lu:2006fr}. }
\begin{center}
\begin{tabular}{l c c c}
\hline
\hline
~~~~~~~~~~~~ &~~~~~~~~~ $a_0(1450)$~~~~~~~~~~~~ & ~~~~~~~~~~~~$K^*_0(1430)$~~~~~~~~~~~~ & ~~~~~~~~~~~~$f_0(1500)$~~~~~~~~~~~~ \\
\hline
$\bar f~{\rm(MeV)}$ &  $460 \pm 50$ & $445 \pm 50$ & $490 \pm 50$ \\
$B_1$ & $-0.58 \pm 0.12$ & $-0.57 \pm 0.13$ & $-0.48 \pm 0.11$ \\
$B_3$ & $-0.49 \pm 0.15$ & $-0.42 \pm 0.22$ & $-0.37 \pm 0.20$ \\
$a_1 (\times 10^{-2})$ & 0 & $1.8 \sim 4.2$ & 0 \\
$a_2$  & $-0.33 \sim -0.18 $ & $-0.33 \sim -0.025$ & $-0.33 \sim -0.18$\\
$a_4$  & $-0.11 \sim 0.39$ & -- & $0.28 \sim 0.79$ \\
$b_1 (\times 10^{-2})$ & 0 & $3.7 \sim 5.5$ & 0 \\
$b_2$ & $0 \sim 0.058$ & $0 \sim 0.15$ &  $-0.15 \sim -0.088$ \\
$b_4$ & $0.070 \sim 0.20$& -- & $0.044 \sim 0.16$\\
\hline
\hline
\end{tabular}
\end{center}\label{results1}
\end{table}

\subsection{Inputs and Light-Cone Distribution Amplitudes}

In the numerical calculation \footnote{Since the helicity form factors for ($B_{(s)} \to S$) (where $S = a_0(1450), K_0^*(1430), f_0(1500)$) in this work are calculated within the two-quark model framework, the LCDA are consequently constructed based on the conventional ($q\bar{q}$) state configuration. We note that possible tetraquark ($qq \bar q \bar q$) or molecular state components in scalar mesons could potentially modify the LCDA parameters and thereby affect the form factors. However, such investigations lie beyond the scope of the current study and will be addressed in our future work.}, we take the input parameters from the review of particle physics (RPP) as follows~\cite{Workman:2022ynf}: $m_{a_0} =1.474 \pm 0.019 \rm{GeV}$, $m_{f_0} =1.506 \pm 0.006 \rm{GeV}$, $m_{K^*_0} =1.425 \pm 0.050 \rm{GeV}$, $m_{B} =5.279 \pm 0.012 \rm{GeV}$, $m_{B_s} =5.367 \pm 0.010 \rm{GeV}$.  The factorization scale $\mu$ is set at the typical momentum transfer of the $B_{(s)} \to S$ decays, {\it i.e.}, $\mu \simeq (m^2_{B_{(s)}}-m^2_b)^{1/2}$~\cite{Ball:2004rg} to effectively separate long-distance (non-perturbative effects) from short-distance (perturbative effects) physics. All numerical calculations presented in this work are performed within the $\overline{MS}$-scheme. And we take the $\overline{MS}$ mass $m_b(m_b)=4.18 \pm 0.05\rm{GeV}$ \cite{Workman:2022ynf}. The renormalization group equation for the $\overline{MS}$ mass can be found in Ref.~\cite{Wang:2014vra}.

Based on the usual correlation function, we need to consider the twist-2 and twist-3 distribution amplitudes of scalar mesons $\phi_{2;S}( u,\mu) $, $\psi_{3;S}^{s}(u,\mu)$ and $\psi_{3;S}^{\sigma}(u, \mu)$. Given the conformal symmetry of the QCD, the twist-2, 3 distribution amplitudes of the scalar mesons, $\phi_{2;S}( u,\mu)$, $\psi_{3;S}^{s}(u,\mu)$ and $\psi_{3;S}^{\sigma}(u, \mu)$, can be expressed by the Gegenbauer polynomials with increasing conformal spin. Furthermore, we shall adopt the model of Ref. \cite{Cheng:2005nb, Lu:2006fr} as follows:
\begin{eqnarray}
\phi _{2;S}\left( u,\mu \right) &=&\bar{f}_{S}\left( \mu \right) 6u\bar{u}%
\left[ B_{0}\left( \mu \right) +\sum_{m=1}^{\infty }B_{m}\left( \mu
\right)
C_{m}^{3/2}\left( 2u-1\right) \right],  \nonumber \\
\psi _{3;S}^{s}\left( u,\mu \right) &=&\bar{f}_{S}\left( \mu \right) \left[
1+\sum_{m=1}^{\infty }a_{m}\left( \mu \right) C_{m}^{1/2}\left( 2u-1\right) %
\right] , \label{Gegexpansion} \\
\psi _{3;S}^{\sigma }\left( u,\mu \right) &=&\bar{f}_{S}\left( \mu \right) 6u%
\bar{u}\left[ 1+\sum_{m=1}^{\infty }b_{m}\left( \mu \right)
C_{m}^{3/2}\left( 2u-1\right) \right],  \nonumber
\end{eqnarray}
where $C_{m}^{1/2}(x)$ and $C_{m}^{3/2}(x)$ are the Gegenbauer polynomials. Moreover, the decay constants $\bar f_S$ of scalar mesons and various Gengenbauer moments $B_m$, $a_m$ and $b_m$ for twist-2 and twist-3 LCDAs have been computed in Refs. \cite{Cheng:2005nb,Lu:2006fr} based on the QCD SVZ sum rules, which are collected in Table \ref{results1}. These parameters can be run into any other scales by using the renormalization group equations.
\begin{figure}[t]
\begin{center}
\includegraphics[width=0.95\textwidth]{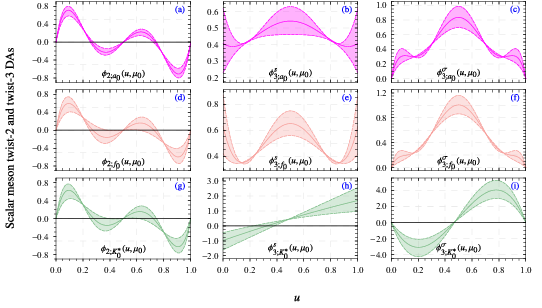}
\end{center}
\caption{Twist-2 and 3 distribution amplitudes of $a_0(1450), f_0(1500), K^*_0(1430)$-mesons at the scale $\mu$ =1~\rm{GeV}. The solid lines are the central values, and the shaded bands are used for the corresponding errors.}
\label{fig1}
\end{figure}
The scale dependence of the decay constants $\bar f_S$ is given as
\begin{eqnarray}
\bar f_S (M)=\bar f_S(\mu)\Bigg( \frac{\alpha_s(\mu)}{\alpha_s(M)} \Bigg)^{4/b},
\end{eqnarray}
For the twist-2,3 distribution amplitude of the scalar mesons, the renormalization group equations of Gegenbauer moments are given as follows:
\begin{eqnarray}
B_m(\mu) &=& B_m(\mu_0)\left(\frac{\alpha_s(\mu_0)}{\alpha_s(\mu)}\right)^{-(\gamma_{(l)}+4)/{b}},\\
\langle a_n (\mu)\rangle&=&\langle a_n (\mu_0)\rangle\bigg({\alpha_s(\mu_0) \alpha_s(\mu)}\bigg)^{-\gamma^S_n /b}, \\
\qquad \langle b_n (\mu)\rangle&=&\langle b_n(\mu_0)\rangle\bigg({\alpha_s(\mu_0) \alpha_s(\mu)}\bigg)^{-\gamma^T_n /b},
\end{eqnarray}
where the one-loop anomalous dimensions are~\cite{Gross:1974cs, Shifman:1980dk}
\begin{eqnarray}
&&\gamma_{(m)} = C_F
\left(1-\frac{2}{(m+1)(m+2)}+4 \sum_{j=2}^{m+1} \frac{1}{j}\right),\\
&&\gamma^S_n = C_F \bigg(1- {8 \over (n+1)(n+2)}+4 \sum_{j=2}^{n+1} \frac{1}{j} \bigg), \\
&&\gamma^T_n = C_F \bigg(1+4 \sum_{j=2}^{n+2} \frac{1}{j}\bigg),
\end{eqnarray}
with $C_F=(N_c^2-1)/(2N_c )=4/3$. Figure~\ref{fig1} shows the twist-2 and twist-3 distribution amplitudes of the scalar mesons with the typical input parameters exhibited in Table~\ref{results1} at the scale $\mu=1\rm{GeV}$. The solid and the shaded bands are for the parameters given in Table~\ref{results1}.

\begin{table}[tb]
\footnotesize
\caption{Central values and uncertainties of the fitting parameters for the HFFs. The upper and lower limits of the parameters correspond to the upper and lower error bands of the HFFs, respectively.}
\begin{center}
\begin{tabular}{c      c      c      c      c      c}
\hline\hline
 HFFs   &   $~~~~~~~~$   &   $~~B \to a_0(1450)~~$   & ~~$B_s \to f_0(1500)~~$   & ~~$B \to K^*_0(1430)$~~   &   $~~B_s \to K^*_0(1430)~~$  \\
\hline                     &  $\alpha_0^{V,0}$  &  $0.28^{+0.04}_{-0.05}$  &  $0.27^{+0.04}_{-0.04}$  &  $0.23^{+0.04}_{-0.04}$  & $0.25^{+0.04}_{-0.04}$  \\
$\mathcal{S}_{V,0}(q^2)$   &  $\alpha_1^{V,0}$  &  $-1.42^{-0.19}_{+0.21}$  &  $-1.77^{-0.21}_{+0.23}$  &  $-0.95^{-0.11}_{+0.12}$  & $-1.17^{-0.14}_{+0.16}$  \\
                           &  $\alpha_2^{V,0}$  &  $6.28^{+1.22}_{-1.30}$ &  $7.80^{+1.5}_{-1.55}$ &  $3.03^{+0.94}_{-0.94}$ & $3.87^{+1.07}_{-1.06}$  \\
\hline                     &  $\alpha_0^{V,t}$  &  $0.10^{+0.01}_{-0.02}$  &  $0.09^{+0.02}_{-0.01}$  &  $0.08^{+0.02}_{-0.01}$  & $0.10^{+0.01}_{-0.02}$  \\
$\mathcal{S}_{V,t}(q^2)$   &  $\alpha_1^{V,t}$  &  $-0.46^{-0.07}_{+0.08}$  &  $-0.57^{-0.08}_{+0.08}$  &  $-0.39^{+0.07}_{-0.06}$  & $-0.50^{-0.08}_{+0.08}$  \\
                           &  $\alpha_2^{V,t}$  &  $2.81^{+0.67}_{-0.70}$ &  $3.20^{+0.75}_{-0.73}$ &  $2.38^{+0.74}_{-0.71}$ & $2.89^{+0.82}_{-0.78}$  \\
\hline                     &  $\alpha_0^{T,0}$  &  $0.29^{+0.05}_{-0.05}$  &  $0.28^{+0.04}_{-0.04}$  &  $0.23^{+0.03}_{-0.04}$  & $0.25^{+0.04}_{0.04}$  \\
$\mathcal{S}_{T,0}(q^2)$   &  $\alpha_1^{T,0}$  &  $-1.00^{-0.10}_{+0.13}$  &  $-1.42^{-0.16}_{+0.17}$  &  $-0.51^{-0.04}_{+0.06}$  & $-0.69^{-0.08}_{+0.09}$  \\
                           &  $\alpha_2^{T,0}$  &  $5.69^{+1.46}_{-1.50}$ &  $6.22^{+1.86}_{-1.87}$ &  $2.93^{+1.42}_{-1.38}$ & $3.74^{+1.56}_{-1.51}$  \\
\hline\hline
\end{tabular}
\end{center}
\label{results3}
\end{table}

\begin{figure}[t]
\begin{center}
\includegraphics[width=0.95\textwidth]{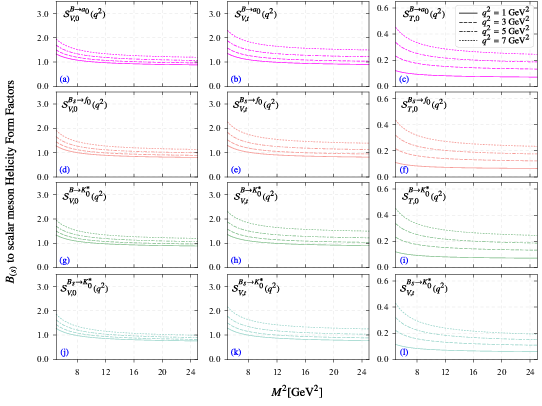}
\end{center}
\caption{The center values for $B_{(s)} \to a_0(1450), f_0(1500), K^*_0(1430)$ HFFs with respect to Borel parameter for the four different $q^2$ values.}
\label{fig6}
\end{figure}

\begin{figure}[t]
\begin{center}
\includegraphics[width=0.95\textwidth]{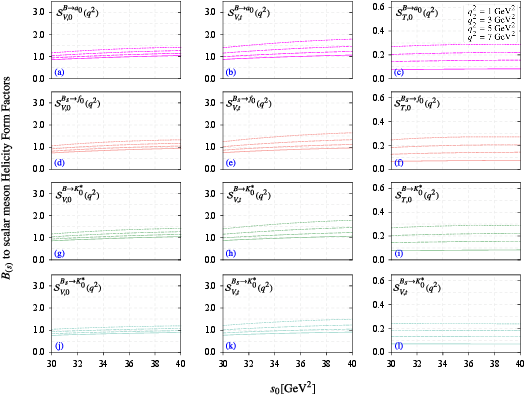}
\end{center}
\caption{The variation of the HFFs for the semi-leptonic transitions $B_{(s)} \to a_0(1450), f_0(1500), K^*_0(1430)$ with the threshold parameter $s_0$ at a fixed value of the Borel parameter $M^2$.}
\label{fig6-1}
\end{figure}

\subsection{The $B_{(s)} \to S$ HFFs}

Using the usual correlation functions, the LCSRs of the $B_{(s)} \to S$ HFFs $\mathcal{S}_{V,0}(q^2)$, $\mathcal{S}_{V,t}(q^2)$ and $\mathcal{S}_{T,0}(q^2)$ can be represented as complex formula containing the twist-2, 3, \dots, distribution amplitudes of the scalar mesons. In this article, we calculate the HFFs up to the twist-3 accuracy and also consider the NLO contributions. Then, we apply the distribution amplitudes discussed above to calculate the HFFs $\mathcal{S}_{V,0}(q^2), ~\mathcal{S}_{V,t}(q^2)$ and $\mathcal{S}_{T,0}(q^2)$~\footnote{While we acknowledge their possible role in certain kinematic regimes, we note that the potential impact of higher-twist contributions are typically associated with large uncertainties, and their contributions are expected to be negligible in our current analysis, as supported by Ref.~\cite{Duplancic:2008tk}. Given the small expected impact and the significant theoretical uncertainties in modeling higher-twist effects, we have chosen to focus on the twist-2,3 contributions in this work.}.

For each scalar meson $S$, there exist three twist-3 distribution amplitudes: the scalar component  $\phi^{s}_{S}$, the tensor component $\phi^{\sigma}_{S}$ and the three-particle component $\phi_{3S}$. The three-particle twist-3 Light Cone Distribution Amplitude can be formally defined through the following matrix element:
\begin{eqnarray}
\langle 0 | \bar{q_2}(x) \sigma_{\mu\nu} g G_{\alpha\beta}(-v x) q_{1}(-x)| S(q)\rangle &=& i \bar{f}_{3S} \big[q_\alpha(q_\mu\delta_{\nu\beta}-q_\nu\delta_{\mu\beta}) - (\alpha\leftrightarrow \beta)\big]\nonumber \\
&\times&\int {\cal D}\alpha_i\, e^{i qx(-\alpha_1+\alpha_2+v\alpha_3)} \phi_{3S}(\alpha_i)
\end{eqnarray}
where ${\cal D}\alpha_i  =d\alpha_1d\alpha_2d\alpha_3\delta(\alpha_1+\alpha_2+\alpha_3-1)$. In the case of the $a_0$-meson, we introduce the parameter $R_1=\frac{1}{m_{a_0}}\frac{\bar{f}_{3a_0}}{\bar{f}_{a_0}}$, where $\bar{f}{3a_0}$ and $\bar{f}_{a_0}$ are decay constants associated with three-particle and local operators respectively. Applying the constraints from the equations of motion and the conformal symmetry for the moments among the twist-3 LCDAs, we derive:
\begin{eqnarray}
\langle\xi^2_{\sigma,a_0}\rangle &=& \frac{1}{5} \langle\xi^0_{\sigma,a_0}\rangle +\frac{12}{5}R_{1}-\frac{8}{5}R_{1}\langle\langle\alpha_3\rangle\rangle_{a_0} \\
\langle\xi^2_{s,a_0}\rangle &=& \frac{1}{3}\langle\xi^0_{s,a_0}\rangle +4R_1
\end{eqnarray}
where the $n$-th moment of the three-particle twist-3 LCDA is defined as $\langle \!\langle(\alpha_2 -\alpha_1+v\alpha_3)^n\rangle \!\rangle
=\int {\cal D}\alpha_{i}\phi_{3 S}(\alpha_i)(\alpha_2 -\alpha_1+v \alpha_3)^n$. Employing the central values for the two particle twist-3 LCDA moments from Ref.~\cite{Han:2013zg}, we obtain
\begin{eqnarray}
\langle\langle\alpha_3\rangle\rangle_{a_0}\sim 1.29 \;\;{\rm and}\;\; f_{3a_0} \sim 4.82\times10^{-3} {\rm GeV}^2 \ ,
\end{eqnarray}
The numerical result ($R_1 \equiv \frac{\bar{f}{3a_0}}{m{a_0}\bar{f}_{a_0}} < 0.01$) reveals that three-particle twist-3 LCDA contributions are suppressed by at least two orders of magnitude relative to two-particle effects. Consequently, the three-particle component can be safely neglected in conventional applications. For a more detailed discussion, please refer to Ref.~\cite{Han:2013zg}. The present work neglects the three-particle distribution amplitude contributions.

\begin{figure}[t]
\begin{center}
\includegraphics[width=0.95\textwidth]{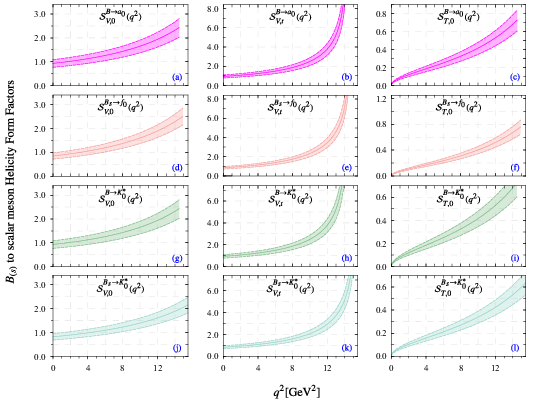}
\end{center}
\caption{HFFs for the semi-leptonic transitions $B_{(s)} \to a_0(1450), f_0(1500), K^*_0(1430)$ across the entire physical region. The solid lines represent the central values, and the shaded bands indicate the corresponding uncertainties obtained by varying the input parameters such as $m_b,~m_{B_{(s)}},~f_{B_{(s)}}$, $m_S,~M^2,~s_0,~\phi_{2;S}(u,\mu),~\psi_{3;S}^{s}(u,\mu),~\psi_{3;S}^{\sigma}(u,\mu),\cdots$, which have been added up in quadrature.}
\label{fig2}
\end{figure}

As for the LCSRs of the HFFs, we need to know the continuum threshold $s_0$ and the Borel parameter $M^2$. For the choices of the Borel parameter $M^2$, On the one hand, the value of $M^2$ must be sufficiently small to ensure that contributions from excited and continuum states are exponentially suppressed. On the other hand, the scale $M^2$ must be sufficiently large to ensure that higher-dimensional operators are effectively suppressed. And considering the stability of $M^2$ for the sum rule, we take it as $12 \pm 1~ \rm{GeV^2}$. To provide intuitive demonstration of the stability criterion, we plot the central values of the form factors versus the Borel mass parameter $M^2$ in Fig.~\ref{fig6}. The figure reveals excellent stability of the form factors with respect to variations in $M^2$ within our chosen windows. The continuum threshold $s_0$ is intrinsically set at the mass of the first excited state of the $B_{(s)}$ mesons ensuring a proper separation between the ground-state pole and the continuum. Besides, the dependence of the HFFs for the semi-leptonic transitions $B_{(s)} \to a_0(1450), f_0(1500), K^*_0(1430)$ on the continuum threshold parameter $s_0$ is also a vital test for the stability and reliability of the QCD sum rules calculation. The Fig.~\ref{fig6-1} showing the variation of the HFFs with the threshold parameter $s_0$ at $M^2=12~\rm{GeV^2}$. As clearly demonstrated in the Fig.~\ref{fig6-1}, the HFFs exhibit a remarkable stability plateau within the range of $s_0\in [30,40]~\rm{GeV^2}$. The variation of the form factor values within this window indicates that our results are insensitive to the precise value of $s_0$ within this region. Based on this, we take $s^B_0=35 \pm 1 ~ \rm{GeV^2}$  and $s_0^{B_s}=36 \pm 1~ \rm{GeV^2}$ to minimize any potential systematic uncertainties. As the LCSRs of the form factors are only reliable in low and intermediate $q^2$ regions, it is essential to employ a form factor parametrization that takes into account the characteristic features of the form factors. For each considered form factor, we define a parametrization based on the series expansion as follows~\cite{Bourrely:2008za,Arnesen:2005ez,Boyd:1994tt,Boyd:1997qw,Becher:2005bg},
\begin{eqnarray}
{\mathcal S}_{V,0}(q^2)  &=& \frac1{B(q^2) \phi_T^V(q^2)} \sum_{k=0}^{K-1} \alpha^{(V,0)}_k z^k, \\
{\mathcal S}_{V,t}(q^2)  &=& \frac1{B(q^2) \sqrt{z(q^2,t_-)} \phi_L^V(q^2)} \sum_{k=0}^{K-1} \alpha^{(V,t)}_k z^k \\
{\mathcal S}_{T,0}(q^2) &=& \frac{\sqrt{-z(q^2,0)}}{B(q^2) \phi_T^T(q^2)} \sum_{k=0}^{K-1} \alpha^{(T,0)}_k z^k
\end{eqnarray}
where
\begin{eqnarray}
\phi_I^X(q^2)&=&\sqrt{\frac{1}{48\pi\chi^{X}_{I}(n)}}\, \frac{(q^2-t_+)}{(t_+-t_0)^{1/4}}
\left(\frac{z(q^2,0)}{-q^2}\right)^{(3+n)/2}\left(\frac{z(q^2,t_0)}{t_0-q^2}\right)^{-1/2}
\left(\frac{z(q^2,t_-)}{t_--q^2}\right)^{-3/4}\,\nonumber\\
z &\equiv& z(q^2,t_0)=\frac{\sqrt{t_+ - q^2}-\sqrt{t_+ - t_0}}{\sqrt{t_+ - q^2}+\sqrt{t_+ - t_0}} \nonumber \\
t_\pm &=& (m_{B_{(s)}} \pm m_S)^2  \\
t_0 &=& t_+(1-\sqrt{1-t_-/t_+}) \nonumber
\end{eqnarray}
The coefficient $\chi^{X}_{I}(n)$ can be calculated by the QCD sum rules. For more specific details, please refer to the references~\cite{Bharucha:2010im,Bourrely:2008za}. The simplified version of the series expansion (SSE) method can be obtained by the following replacements: the Blaschke factor $B(q^2)=1-q^2/m^2_S$, $\phi_I^X(q^2) \to 1$, $\sqrt{-z(q^2,0)} \to \sqrt{q^2}/m_{B_{(s)}}$, and $\sqrt{z(q^2,t_-)} \to \sqrt{\lambda(q^2)}/m_{B_{(s)}}^2$. We present the fitted parameters of the HFFs extrapolation in Table~\ref{results3},. We show the extrapolated HFFs of $B_{(s)} \to S$ in Figure~\ref{fig2}, where the shaded hands are the theoretical uncertainties from all the error sources, such as $m_b,~m_{B_{(s)}},~f_{B_{(s)}},~m_S,~M^2,~s_0,~\phi_{2;S}(u,\mu),~\psi_{3;S}^{s}(u,\mu),~\psi_{3;S}^{\sigma}(u,\mu),\cdots$, which have been added up in quadrature. In Table~\ref{results4}, the LO and NLO values of HFFs $\mathcal{S}_{V,(0,t)}(q^2=1~\rm{GeV^2})$ and $\mathcal{S}_{T,(0)}(q^2=1~\rm{GeV^2})$ of $B_{(s)} \to S$ transitions are presented. All the NLO corrections are negative and amount at most to about $25\%$.

\begin{table}[t]
\footnotesize
\caption{The values of the HFFs for the transitions $B \to a_0(1450)$, $B_s \to f_0(1500)$, $B \to K^*_0(1430)$ and $B_s \to K^*_0(1430)$ from the LCSRs at the point of $q^2=1~\rm{GeV^2}$.}
\begin{center}
\begin{tabular}{c  c  c  c  c  c}
\hline\hline
 HFFs   &   $~~~~~~$   &      ~$B \to a_0(1450)$~&   ~$B_s \to f_0(1500)$~   &   ~$B \to K^*_0(1430)$~   &   ~$B_s \to K^*_0(1430)$~  \\
\hline                &  Total  &  $0.97$  &  $0.89$  &  $0.79$  & $0.86$  \\
$\mathcal{S}_{V,0}~(q^2=1~\rm{GeV^2})$   &  LO         &  $1.13$  &  $1.01$  &  $0.92$  & $0.99$  \\
                      &  NLO        &  $-0.16$ &  $-0.12$ &  $-0.13$ & $-0.13$  \\
\hline                &  Total  &  $0.99$  &  $0.90$  &  $0.80$  & $0.87$  \\
$\mathcal{S}_{V,t}~(q^2=1~\rm{GeV^2})$   &  LO         &  $1.15$  &  $1.02$  &  $0.93$  & $1.00$  \\
                      &  NLO        &  $-0.16$ &  $-0.12$ &  $-0.13$ & $-0.13$  \\
\hline                &  Total  &  $0.08$  &  $0.07$  &  $0.07$  & $0.07$  \\
$\mathcal{S}_{T,0}~(q^2=1~\rm{GeV^2})$   &  LO         &  $0.10$  &  $0.09$  &  $0.08$  & $0.08$  \\
                      &  NLO        &  $-0.02$ &  $-0.01$ &  $-0.01$ & $-0.01$  \\
\hline\hline
\end{tabular}
\end{center}
\label{results4}
\end{table}

\subsection{Applications of the $B_{(s)} \to S$ HFFs}

Within the derived HFFs, one can calculate some phenomenologically significant observables, such as the differential decay widths, the branching ratio, and the polarization asymmetry. Next, as an application of these HFFs, we discuss the properties of the semi-leptonic channels, $B_{(s)} \to S \ell \bar\nu_\ell$ and $B_{(s)} \to S \ell \bar\ell/\nu \bar\nu$. The differential decay widths can be expressed as follows~\cite{Sun:2010nv, Khosravi:2022fzo, Han:2013zg, Yang:2005bv},
\begin{align}
& \frac{d\Gamma}{dq^2}(B_{(s)}\to S \ell \bar{\nu}_\ell) = \frac{G_F^2|V_{ub}|^2}{192 \pi^3 m_{B_{(s)}}^3} \frac{q^2 \! - \! m_\ell^2}{q^4} \sqrt{ \frac{(q^2  \!  -  \!  m_\ell^2)^2}{q^2}} \sqrt{\frac{(m_{B_{(s)}}^2 \! - \! m_S^2 \! - \! q^2)^2} {4q^2}-m_S^2} \nonumber\\
&\hspace{1.6cm}\times \bigg\{\frac14  (m_\ell^2  +  2q^2) (q^2  -  (m_{B_{(s)}}  -  m_S)^2)(q^2  -  (m_{B_{(s)}}  +  m_S)^2)|\mathcal{S}_{V,0}(q^2)|^2 \nonumber\\
&\hspace{1.6cm}+ 3m_\ell^2(m_{B_{(s)}}^2-m_S^2)^2 \Big\{\frac12\mathcal{S}_{V,0}(q^2)+ \frac{q^2} {m_{B_{(s)}}^2-m_S^2}  \Big[\frac12 \mathcal{S}_{V,0}(q^2)- \frac1{2 q^2}\Big( \sqrt{\lambda} \nonumber\\
&\hspace{1.6cm}\times \mathcal{S}_{V,t}(q^2) - (m_{B_{s}}^2 - m_S^2-q^2)\mathcal{S}_{V,0}(q^2)\Big)\Big] \Big\}^2\bigg\}
\label{dgamma1}
\end{align}
\begin{eqnarray}
\frac{d\Gamma}{dq^2}(B_{(s)}\to S\ell^+\ell^-) &=& \frac{G_F^2 | V_{tb}V_{ts}^*| ^2 \, m_{B_{(s)}}^{3}  \,  \alpha _{\rm em}^2 }{1536 \pi ^{5}}  \, \Big( 1-\frac{4r_\ell}{s}\Big) ^{1/2} \left[ \left( 1+\frac{2r_\ell}{s}\right)\varphi _{S}^{3/2} \alpha_{S} \right.\nonumber\\
\hspace{1.6cm} &+& \left. \varphi _{S}^{1/2}r_\ell\delta _{S}\right]
\label{dgamma2}
\\[2ex]
\frac{d{\Gamma}}{dq^2}(B_s\rightarrow K_0^*(1430) {\nu} \bar \nu) &=& \frac{ G_{F}^2 |V_{td}V^*_{tb}|^2 m_{B_s}^3  \alpha^2} {2^8 \pi ^5} \frac{{\mid}D_{\nu}(x_t){\mid}^2}{\sin^4\theta_W} \phi ^{3/2}(1,\hat{r},\hat{s}) \Big|\frac12 \mathcal{S}_{V,0}(q^2)\Big|^2
 \label{dgamma3}
\end{eqnarray}
where
\begin{align}
& s =\frac{q^2}{m_{B_{(s)}}^2} , \; r_\ell=\frac{m_\ell^2}{m_{B_{(s)}}^2} ,\; r_{S}=\frac{m_{S}^2}{m_{B_{(s)}}^2} ,~x_t = \frac{m_t^2}{m_W^2}
\\
&\varphi_S =(1-r_S) ^2 - 2s(1+r_S) +s^2 ,
\\
&\alpha_S = \Big|\frac12 C_9^{\rm eff}{\mathcal{S}_{V,0} (q^2) }
-2 \frac{m_{B_s} - m_S}{q^2} \frac{C_7}{1+\sqrt{r_S}} \mathcal{S}_{T,0}(q^2)\Big|^2+\Big|\frac12C_{10}{\mathcal{S}_{V,0}(q^2) }\Big|^2 ,
\\
& \delta_S= 6 \, |C_{10}|^2 \, \bigg\{[ 2(1\,+\,r_S) \,-\,s] \Big|\frac12\mathcal{S}_{V,0}(q^2) \Big|^2 \,+\, (1 \,- \, r_S) 2{\rm Re} \bigg\{\frac12 \mathcal{S}_{V,0}\, (q^2 )
\nonumber\\
&\qquad\times \Big[\frac12 \mathcal{S}_{V,0}(q^2) \,-\, \frac1{2 q^2} \Big(\sqrt{\lambda} \mathcal{S}_{V,t}(q^2) \,- (m_{B_{s}}^2 \,- m_S^2-q^2)\mathcal{S}_{V,0}(q^2)\Big)\Big]^\ast \Big\}
\nonumber\\
&\qquad +s \Big| \frac12 \mathcal{S}_{V,0}(q^2) \,-\,  \frac1{2 q^2} \Big(\sqrt{\lambda} \mathcal{S}_{V,t}(q^2) \,- (m_{B_{s}}^2 - m_S^2-q^2\Big)\mathcal{S}_{V,0}(q^2)) \Big|^2 \bigg\}
\\
&D_{\nu}(x_t)= \frac{x_t}{8} \bigg[\frac{2+x_t}{x_t-1}+\frac{3x_t-6}{(x_t-1)^2}\ln x_t\bigg]
\end{align}
In the numerical calculation, we take the CKM matrix elements $|V_{td}V^*_{tb}|=0.008$, $|V_{tb}|=0.9991$, $|V_{ts}|=41.61^{+0.10}_{-0.80} \times 10^{-3}$ and $|V_{ub}|= (2.84\pm0.05) \times 10^{-3}$ \cite{Zhong:2022ecl,Khosravi:2022fzo,Wang:2008da}, the Fermi coupling constant $G_F=1.166 \times 10^{-5} \rm{GeV^{-2}}$, $m_\ell$ denotes the mass of a final-state lepton. These expressions also contain the Wilson coefficients $C^{\rm{eff}}_7=-0.313$, $C_9^{\rm{eff}}$ and $C_{10}=-4.669$, the form factors related to the DAs, and decay constants~\cite{Buras:1994dj, Aliev:2005jn, Misiak:1992bc, Buchalla:1995vs, Hatanaka:2008gu}. Except for the $B_{(s)} \to S$ HFFs, we adopt the other input parameters as those of Ref.~\cite{Sun:2010nv}.

\begin{figure}[t]
\begin{center}
\includegraphics[width=0.9\textwidth]{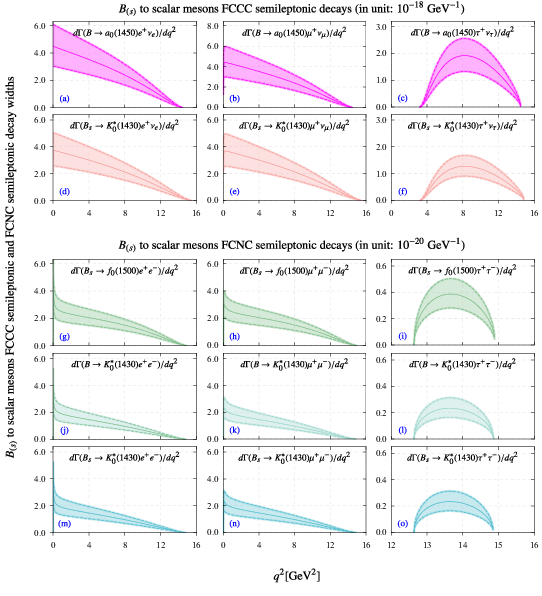}
\end{center}
\caption{The differential decay widths for the FCCC semi-leptonic decays $B \to a_0(1450)\ell\bar\nu_\ell$, $B_s \to K_0^*(1430)\ell\bar\nu_\ell$, FCNC semi-leptonic decays $B_s \to f_0(1500)\ell^+\ell^-$, $B_{(s)} \to K_0^*(1430)\ell^+\ell^-$, which the lepton are taken as $e,\mu, \tau$. The solid lines are the central values, and the shaded bands are used for the corresponding errors.}
\label{fig3}
\end{figure}

To be more specific, the effective Wilson coefficient $C_9^{eff}$ can decompose as follows:
\begin{eqnarray}
C_9^{\rm{eff}}(\mu)&=&C_9(\mu)+Y_{\rm{SD}}(z,s')+Y_{\rm{LD}}(z,s'),
\end{eqnarray}
where $Y_{\rm{SD}}(z,s')$ and $Y_{\rm{LD}}(z,s')$ stand for the short-and long-distance contributions from the four quark operators respectively\cite{Buras:1994dj}. The former can be calculated using the perturbative theory, while the latter requires the first-principles QCD computations. With
\begin{eqnarray}
Y_{\rm{SD}}(z,s') &=& h(z,s')C_0-\frac{1}{2}h(1,s')(4C_3+4 C_4+3C_5+C_6)\nonumber\\
&&- \frac{1}{2}h(0,s')(C_3+3C_4) + \frac{2}{9}(3C_3 + C_4 +3C_5+ C_6)
\end{eqnarray}
where
\begin{eqnarray}
C_0 &=& 3C_1+C_2+3C_3+C_4+3C_5 +C_6\\
h(z,s') &=& -{8\over 9}{\rm {ln}}z+{8\over 27}+{4\over 9}x-{2\over
9}(2+x)|1-x|^{1/2}
 \left\{
\begin{array}{l}
\ln \left| \frac{\sqrt{1-x}+1}{\sqrt{1-x}-1}\right| -i\pi \quad {\rm {for}}%
{ {\ }x\equiv 4z^{2}/s^{\prime }<1} \\
2\arctan \frac{1}{\sqrt{x-1}}\qquad {\rm {for}}{ {\ }x\equiv
4z^{2}/s^{\prime }>1}
\end{array}
\right., \nonumber\\
h(0,s^{\prime}) &=& {8 \over 27}-{8 \over 9} {\rm ln}{m_b \over \mu}
-{4 \over 9} {\rm {ln}}s^{\prime} +{4 \over 9}i \pi \,\, ,
\end{eqnarray}
and $z=m_c/m_b$ and $s^{\prime}=q^2/m^2_b$.

For further technical details and supporting literature, we refer the reader to the Ref~\cite{Buras:1994dj,Aliev:2005jn,He:1988tf,Grinstein:1988me,Deshpande:1988bd,ODonnell:1991cdx,Paver:1991tn,
Buchalla:1995vs,Ali:1991is}.

Our prediction for the differential decay widths are presented in Figure~\ref{fig3} \footnote{It should be noted that the light-cone sum rules (LCSRs) have inherent limitations in modeling resonant effects. Consequently, the predictions in the charmonium resonance regions of this figure are subject to corresponding theoretical uncertainties stemming from this approximation.} and Figure~\ref{fig4}, which have a different arising trend versus the $q^2$ and the uncertainties from all error sources are added in quadrature. Furthermore, integrating Eqs.~\eqref{dgamma1}-\eqref{dgamma3} over $q^2$ in the whole physical region $m_\ell ^2 \leq q^2 \leq {(m_{B_{(s)}}- m_S)}^2$ and $4 m_\ell ^2 \leq q^2 \leq {(m_{B_s}- m_{K_0^*})}^2$, and using the total mean lifetime $\tau_{B_{s}}$ \cite{Workman:2022ynf}, we present the branching ratios of the semi-leptonic decays $B_{s}\to K_0^*(1430)\, \ell\, \bar{\nu}_\ell $,  ($\ell=\mu, \tau$) and $B_s \rightarrow K_0^*(1430) \ell^{+}\ell^{-}/\nu\bar{\nu}$ in Table~\ref{Tab:branching}. Here, we should stress that the results obtained for the electron are very close to the muon results. Table~\ref{Tab:branching} contains the results estimated via the conventional QCDSR~\cite{Yang:2005bv}, PQCD~\cite{Li:2008tk}, LCSRs~\cite{Han:2023pgf,Wang:2008da}, and RQM~\cite{Faustov:2013ima} approaches. Considering the uncertainties, the values obtained in this work agree with the LCSRs and PQCD results. As can be seen in this table, the uncertainties obtained for the branching ratios of the semi-leptonic decays $B_{s}\to K_0^*(1430) \ell \bar{\nu}_\ell$ are very large.

\begin{figure}[t]
\begin{center}
\includegraphics[width=0.5\textwidth]{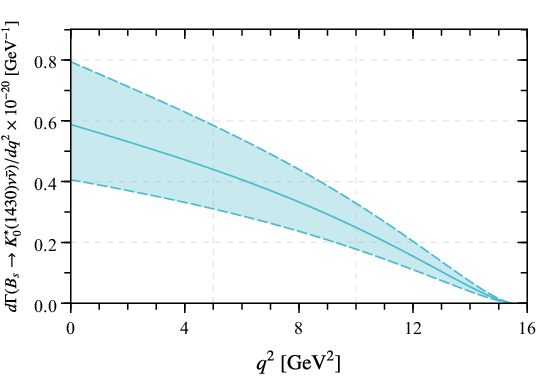}
\end{center}
\caption{Differential decay width for the semileptonic $B_s \to K_0^*(1430) \nu \bar\nu$. The solid lines are the central values, and the shaded bands are used for the corresponding uncertainties.}
\label{fig4}
\end{figure}

\begin{table}[tb]
\footnotesize
\renewcommand{\arraystretch}{1.3}
\begin{center}
\caption{Branching fractions of the FCCC semi-leptonic decays $B_{(s)} \to S\ell\bar\nu_\ell$ and FCNC semi-leptonic decays $B_{(s)} \to S\ell^+\ell^-$. The uncertainties are squared averages of all the mentioned error sources. For comparison, we also present the prediction of various methods.}\label{Tab:branching}
\begin{tabular}{l l l l }
\hline
~~~~~~~~~~~~~~~~~~~~~~&$B^+\to a_0(1450) e^+ \nu_e$~~~~~~~~~~&$B^+ \to a_0(1450) \mu^+ \nu_\mu$~~~~~~~~~~&$B^+ \to a_0(1450) \tau^+ \nu_\tau$\\
This Work & $8.20 ^{+2.80}_{-2.58} \times 10^{-5}$ & $8.17^{+2.79}_{-2.57}\times 10^{-5}$ & $3.36^{+1.13}_{-1.05}\times 10^{-5}$ \\
pQCD~\cite{Li:2008tk}  & $ $ & $3.25^{+2.36}_{-1.36}\times 10^{-4}$ & $1.32^{+0.97}_{-0.57}\times 10^{-4}$ \\
LCSR~\cite{Han:2023pgf}  & $ $ & $1.00(43)\times 10^{-4}$ & $3.0(12)\times 10^{-5}$ \\
LCSR~\cite{Wang:2008da}  & $ $ & $1.8^{+0.9}_{-0.6}\times 10^{-4}$ & $6.3^{+3.4}_{-2.5}\times 10^{-5}$ \\
\hline
& $B_s^0 \to f_0(1500) e^+ e^-$ & $B_s^0 \to f_0(1500) \mu^+ \mu^-$ & $B_s^0\to f_0(1500) \tau^+ \tau^-$ \\
This Work  & $5.09^{+1.57}_{-1.42} \times 10^{-7}$ & $5.00 ^{+1.53}_{-1.39}\times 10^{-7}$ & $1.6^{+0.05}_{-0.04}\times 10^{-6}$ \\
LCSR~\cite{Han:2023pgf}  &  & $2.50(97)\times 10^{-6}$ &  \\
\hline
& $B^+\to K^{*+}_0(1430) e^+ e^-$ & $B^+ \to K^{*+}_0(1430) \mu^+ \mu^-$ & $B^+ \to K^{*+}_0(1430) \tau^+ \tau^-$ \\
This Work  & $3.52 ^{+1.17}_{-1.04} \times 10^{-7}$ & $3.45^{+1.15}_{-1.02}\times 10^{-7}$ & $0.09 ^{+1.96}_{-1.74}\times 10^{-7}$ \\
\hline
& $B_s^0 \to K^{*-}_0(1430) e^+ \nu_e$  &  $B_s^0 \to K^{*-}_0(1430) \mu^+ \nu_\mu$ & $B_s^0 \to K^{*-}_0(1430) \tau^+ \nu_\tau$ \\
This Work & $7.33 ^{+2.43}_{-2.17}\times 10^{-5}$ & $7.31 ^{+2.42}_{-2.16} \times 10^{-5}$ & $3.30 ^{+1.08}_{-0.96}\times 10^{-5}$ \\
QCDSR~\cite{Yang:2005bv} & $ $ & $3.6^{+3.8}_{-2.4}\times 10^{-5}$ & $ $ \\
pQCD~\cite{Li:2008tk} & $0.245^{+0.177}_{-0.105} \times 10^{-4}$ & $0.245^{+0.177}_{-0.105} \times 10^{-4}$ & $0.245^{+0.177}_{-0.105} \times 10^{-4}$ \\
LCSR~\cite{Huang:2022xny} & $0.113^{+0.074}_{-0.051} \times 10^{-4}$ & $0.113^{+0.074}_{-0.051} \times 10^{-4}$ & $0.050^{+0.040}_{-0.020} \times 10^{-4}$ \\
LCSR~\cite{Han:2023pgf} & $ $ & $1.60(67)\times 10^{-4}$ & $5.2(20)\times 10^{-5}$ \\
RQM~\cite{Faustov:2013ima} & $0.71^{+0.14}_{-0.14} \times 10^{-4}$ & $0.71^{+0.14}_{-0.14} \times 10^{-4}$ & $0.21^{+0.14}_{-0.14} \times 10^{-4}$                 \\
LCSR~\cite{Wang:2008da} & $ $ & $1.3^{+1.2}_{-0.4} \times 10^{-4}$ & $5.2^{+5.7}_{-1.8}\times 10^{-5}$ \\
\hline
& $B_s^0 \to K_0^{*0}(1430) e^+ e^-$ & $B_s^0 \to K^{*0}_0(1430) \mu^+ \mu^-$ & $B_s^0 \to K^{*0}_0(1430) \tau^+ \tau^-$ \\
This Work  & $4.70 ^{+1.54}_{-1.38} \times 10^{-7}$ & $4.60 ^{+1.53}_{-1.36}\times 10^{-7}$ & $2.02 ^{+0.256}_{-0.228}\times 10^{-8}$ \\
\hline
& $B_s^0 \to K^{*0}_0(1430) \nu \bar{\nu}$ \\
This Work & $1.15 ^{+0.38}_{-0.34} \times 10^{-7}$ \\
LCSR~\cite{Han:2023pgf} & $2.5(97) \times 10^{-6}$ \\
\hline
\end{tabular}
\end{center}
\end{table}
\begin{figure}[t]
\begin{center}
\includegraphics[width=0.45\textwidth]{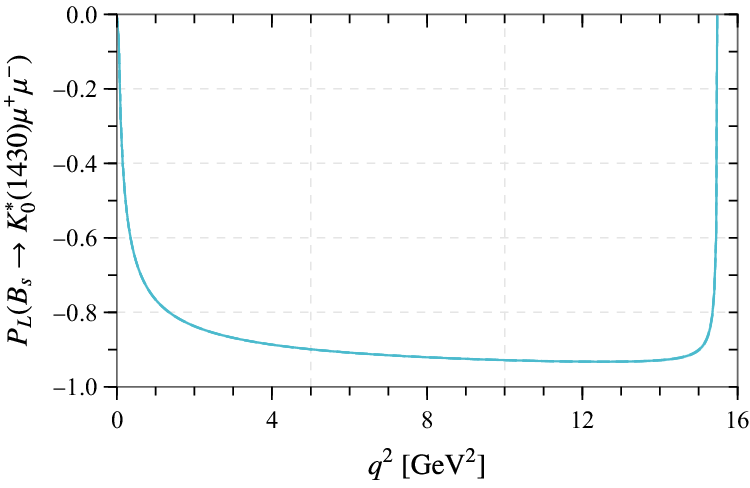}
\includegraphics[width=0.45\textwidth]{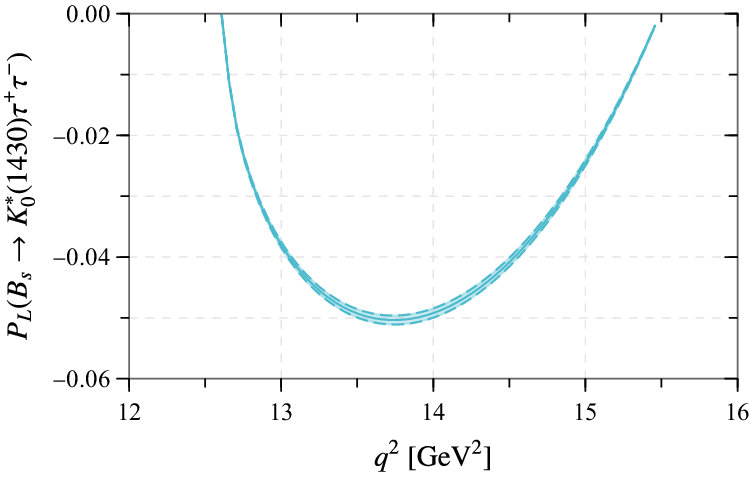}
\includegraphics[width=0.45\textwidth]{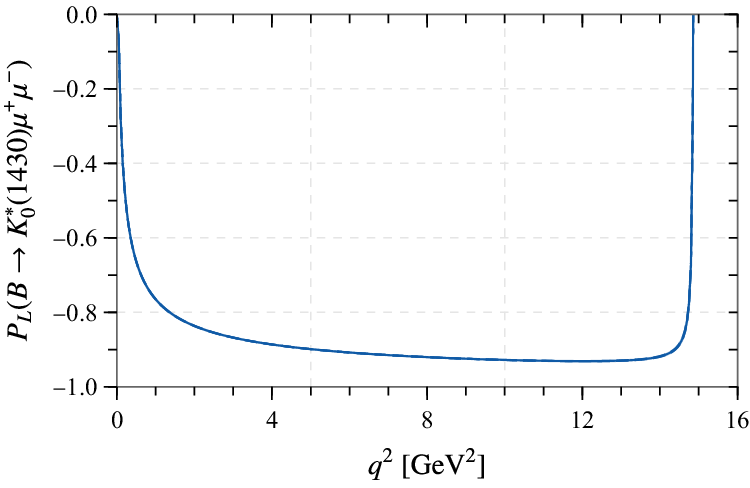}
\includegraphics[width=0.45\textwidth]{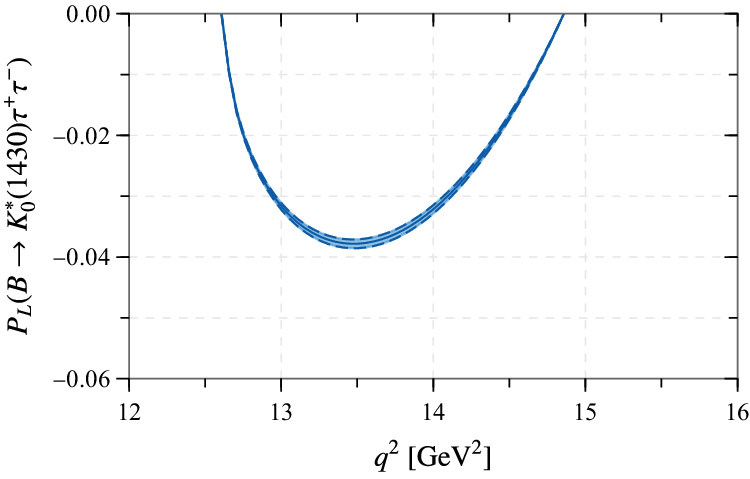}
\end{center}
\caption{Dependence of the longitudinal lepton polarization asymmetries on the squared momentum transfer for the FCNC $B_{(s)} \to K^*_0 (1430 )\ell^+ \ell^-$ decays with $\ell=(\mu,\tau)$. The solid lines are the central values, and the shaded bands are used for the corresponding uncertainties.}
\label{fig5}
\end{figure}
Moreover, the HFFs can also be used for calculating the longitudinal lepton polarization asymmetries $P_L (q^2)$ for the FCNC $B_{(s)} \to K_0^*(1430) \ell^{+} \ell^{-}$ decays. The longitudinal lepton polarization asymmetry for $B_{(s)} \to K_0^*(1430) \ell^{+}\ell^{-}$ can be calculated as follows:

\begin{eqnarray}
P_L (q^2) &=& \frac{2}{(1+\frac{2r_\ell}{s})\phi(1,r,s)\alpha_1+12 r_\ell \beta_1} \sqrt{1-\frac{4r_\ell}{s}} {\rm{Re}} \bigg[\phi(1,r,s) \bigg(-\frac12 C_9^{eff} \mathcal{S}_{V,0} (q^2)   \nonumber \\
& -& \frac{m_{B_{(s)}} + m_S}{q^2} \frac{2C_7 \mathcal{S}_{T,0} (q^2)}{1+\sqrt{r_S}} \bigg)  \bigg(-\frac12 C_{10} \mathcal{S}_{V,0} (q^2)\bigg)^* \bigg],
\end{eqnarray}
where
\begin{align}
\alpha_1 &= \Big|\frac12 C_{9}^{\rm eff} \mathcal{S}_{V,0}(q^2) - \frac{m_{B_{(s)}}+m_S}{q^2} \frac{2\,{\hat m}_b\,C_{7}^{\rm eff} }{1+\sqrt{\hat{r}}} \mathcal{S}_{T,0}(q^2) \Big|^2
+\Big|\frac{1}{2} C_{10} \mathcal{S}_{V,0} (q^2)\Big|^2 ,
\\
\beta_1 &= |C_{10}|^2 \bigg\{ \biggl( 1 \! +\! \hat{r} -\! {\hat{s}\over 2}\biggr) \Big|-\frac12 \mathcal{S}_{V,0}(q^2)\Big|^2 +( 1\! -\! \hat{r}) {\rm Re}\bigg\{-\frac12 \mathcal{S}_{V,0}(q^2) \bigg[-\frac{1}{2}
\nonumber\\
&\times   \mathcal{S}_{V,0}(q^2) \,-\, \frac1{2 q^2} [\sqrt{\lambda} \mathcal{S}_{V,t}(q^2) \,- (m_{B_{(s)}}^2 \,- \, m_S^2-q^2)\mathcal{S}_{V,0}(q^2)]\bigg] ^*\bigg\}
+ \frac12\hat{s}
\nonumber\\
&\times \Big|\frac{1}{2}\, \mathcal{S}_{V,0}(q^2)\,+\, \frac1{2 q^2}\, [\sqrt{\lambda}\, \mathcal{S}_{V,t}(q^2) \,-\, (m_{B_{(s)}}^2 \,-\, m_S^2\,-\,q^2)\,\mathcal{S}_{V,0}(q^2)\,]\,\Big|^2 \,\bigg\}
\end{align}
where $l, r, s$ and $ \phi(1,r,s)$ were defined in Refs.~\cite{Bobeth:1999mk,Mahmoudi:2018qsk}. The dependence of the longitudinal lepton polarization asymmetries for the $B_s \to K_0^*(1430) \ell^{+} \ell^{-}$ decays with $\ell=(\mu, \tau)$ on the transferred momentum square $q^2$ are plotted in Figure~\ref{fig5}. These polarization asymmetries provide valuable information on the flavor-changing loop effects in the SM. As shown in Figure~\ref{fig5}, $P_L(q^2)$ of the semi-leptonic $B_s \to K_0^*(1430) \mu^+ \mu^-$ decay is close to $-1$ except those in the endpoint region and $P_L(q^2)$ of $B_s \to K_0^*(1430) \mu^+ \mu^-$ ranges from $-0.055$ to $0$. For the semi-leptonic $B \to K_0^*(1430) \mu^+ \mu^-$ decay, the $P_L$ is close to $-1$, which in similar to the $B_s \to K_0^*(1430) \mu^+ \mu^-$ decay. The $P_L$ of the $B \to K_0^*(1430) \tau^+ \tau^-$ is range form $-0.038$ to $0$. It is evident that, due to the limited detection efficiency for $\tau$ leptons, measuring the $\tau$ polarization in the near future remains unfeasible.

\section{Summary}\label{sec:4}
In this work, we have performed a systematic analysis of the HFFs for the transitions $B_{(s)} \to S$ $(a_0(1450)$, $K_0^*(1430)$, $f_0(1500))$ within the LCSRs. More precise theoretical predictions have been attempted considering the NLO corrections and keeping the twist-3 LCDAs. We have then adopted a $z$-series expansion to extrapolate the HFFs over the full physical $q^2$-range to obtain results valid for all the relevant kinematic regions for these decays, the behaviors of the HFFs across the entire physical regions have calculated and shown in Figure~\ref{fig2}. At $q^2=1\rm{~GeV^2}$, the NLO corrections are negative for all HFFs, amounting to approximately $25\%$. The results have exhibited in Table~\ref{results4}. This underlines the importance of the NLO contributions to get a more accurate theoretical description of these semi-leptonic decays. In addition, we have used the computed HFFs based on the conventional correlation functions to analyze several physical observables like differential decay widths, branching ratios, and lepton polarization asymmetries for the semi-leptonic processes $B_{(s)} \to S \ell \bar{\nu_\ell}$, $B_{(s)} \to S \ell \bar{\ell}$ and $B_{(s)} \to S \nu \bar{\nu}$. The curves for the semi-leptonic decays have been shown in Figure~\ref{fig3}, Figure~\ref{fig4}, and Figure~\ref{fig5}. From our results of the branching ratios for these decays, one can see that agree with the values reported in pQCD and LCSRs, but differ significantly from the other predictions, which needs more theoretical studies to analyze the quark composition and related properties of these scalar mesons. Given that these decays have not yet been observed experimentally, our theoretical predictions highlight the necessity for future precision experiments to undertake their measurement and establish conclusive verification.

\hspace{2cm}

\acknowledgments
The authors wish to express their gratitude to Xing-Gang Wu and Zhi-Gang Wang for their helpful comments and constructive discussions. LSG is partly supported by the National Key R\&D Program of China under Grant No. 2023YFA1606700 and the National Science Foundation of China under Grant No. 12435007. This work was also supported in part by the China Postdoctoral Science Foundation Grant under No. 2023M740190, the National Natural Science Foundation of China under Grant  No.12265009, No.12265010, No.12347182, the Chongqing Natural Science Foundation project under Grant No. CSTB2022NSCQ-MSX0432, the Science and Technology Research Project of Chongqing Education Commission under Grant No. KJQN202200621 and No. KJQN202200605, the Chongqing Human Resources and Social Security Administration Program under Grants No. D63012022005, the Project of Guizhou Provincial Department of Science and Technology under Grant No.ZK[2025]MS219, No.ZK[2023]024. This work has been supported by the National Natural Science Fund of China (Grant No.12147102).

\end{document}